\documentclass[aps,prl,twocolumn,showpacs,groupedaddress]{revtex4}

\usepackage{color} 
\usepackage{graphicx}
\usepackage{epsfig}
\preprint{LuFeO-2010}
\begin{document}
\DeclareGraphicsExtensions{.eps, .jpg}
\bibliographystyle{prsty}
\input epsf

\title{Infrared study  of the charge-ordered multiferroic  LuFe$_{2}$O$_{4}$} 
\author {F. M. Vitucci$^{1,2}$, A. Nucara$^{1}$, D. Nicoletti$^{1}$, Y. Sun$^{3}$, C. H. Li$^{3}$, J. C. Soret$^{2}$, U. Schade$^{4}$, and P. Calvani$^{1}$}
\affiliation {$^{1}$CNR-SPIN and Dipartimento di Fisica, Universit\`a di Roma La Sapienza,
Piazzale A. Moro 2, I-00185 Roma, Italy} 
\affiliation {$^{2}$LEMA-CNRS, Universit\'e de Tours, F-37200 Tours, France}
\affiliation {$^{3}$Beijing National Laboratory for Condensed Matter Physics, Institute of Physics, Chinese Academy of Sciences, Beijing 100080, P. R. China}
\affiliation{$^{4}$Berliner Elektronenspeicherring-Gesellschaft f\"ur 
Synchrotronstrahlung m.b.H., Albert-Einstein Strasse 15, D-12489 Berlin, 
Germany}
\date{\today}

\begin{abstract}
The reflectivity of a large LuFe$_{2}$O$_{4}$ single crystal has been measured with the radiation field  either perpendicular or parallel to the $c$ axis of its rhombohedral structure, from 10 to 500K, and from 7 to 16000 cm$^{-1}$. The transition  between the two-dimensional and the three-dimensional charge order at $T_{CO}$ = 320 K is found to change dramatically the phonon spectrum  in both polarizations. The number of the observed modes above and below $T_{CO}$, according  to a factor-group analysis, is in good agreement with a transition from the rhombohedral space group $R{\bar 3}m$  to the monoclinic $C2/m$. In the sub-THz region a peak  becomes evident  at low temperature, whose origin is discussed in relation with previous experiments. 
\end{abstract}


\maketitle

\section{Introduction} 

The multiferroics are materials which exhibit, in a common range of temperatures,   both ferroelectricity and magnetic order (ferromagnetism, ferrimagnetism or antiferromagnetism).  Much effort has been  devoted to understand the relationship between their magnetic and electronic degrees of freedom, \cite{Eerenstein,Spaldin,Hill} in view of the potential applications. Indeed, the conventional mechanisms of ferroelectricity and magnetism are  not compatible with each other, and this makes the multiferroic materials  relatively rare. One should then invoke novel mechanisms of ferroelectricity, as those associated with specific charge configurations in charge-ordered materials. \cite{Brink} One of such compounds is the mixed-valence LuFe$_{2}$O$_{4}$ (LFO). Its structure belongs to the rhombohedral space group $R{\bar 3}m$ and includes triangular double layers of Fe-O stacked along the $c$ axis. A two-dimensional Fe$^{2+}$/Fe$^{3+}$ charge order (CO), which  forms between about 500 K and $T_{CO}$ = 320 K in those double layers, builds up in each Fe-O plane a net dipole moment. \cite{Tanaka,Ikeda2,Ikeda} Below $T_{CO}$, the CO becomes more robust and three-dimensional (3-D). 

Dielectric measurements in low electric fields show that the 3-D phase of LFO  is ferroelectric, \cite{Ikeda} while X-ray scattering indicates that the dipole moments of adjacent bilayers are antiparallel, \cite{Angst} so that the net polarization is zero. These contradictory findings can be reconciled by considering that the  antiferroelectric ground state and the ferroelectric state may be so close in energy \cite{Angst} that even a small electric field  stabilizes the latter one. Concerning the magnetic behavior of LFO, a two-dimensional  ferrimagnetic order is established below T$_{N}$ $\simeq$ 240 K \cite{Yamada}. However, a surprising decrease in the magnetic correlation length is observed below T$_{L}$ = 175 K. \cite{Christi}  Finally, it has been found that the dielectric constant decreases sharply both under weak magnetic \cite{Subramanian} and electric \cite{Li2} fields, which also cause a drop in the resistivity by several orders of magnitude.  \cite{Li} This colossal electroresistance, which is also observed at room temperature, makes LFO attractive for its potential applications.

%<<<<<<<<<<<<<<<<<<<<<<<<<<<<<<<< FIG 1 >>>>>>>>>>>>>>>>>>>>>>>>>>>>>>>>>>

\begin{figure}[htbp]   
\begin{center}  
\leavevmode    
\epsfxsize=8.6cm \epsfbox {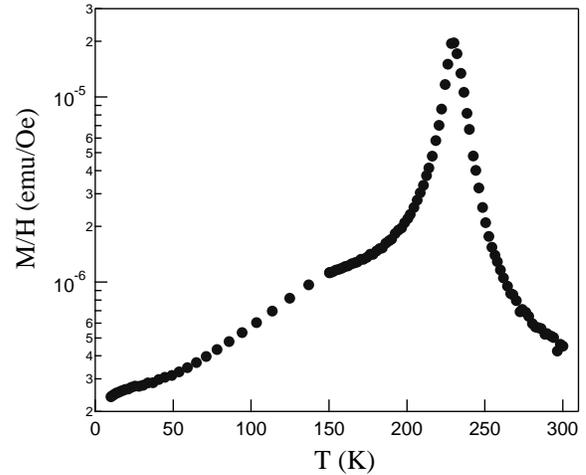} 
\caption{Magnetic susceptibility of the single crystal of LuFe$_{2}$O$_{4}$, as measured in zero-field cooling.}
\label{chi}
\end{center}
\end{figure}
%<<<<<<<<<<<<<<<<<<<<<<<<<<<<<<<< END FIG 1 >>>>>>>>>>>>>>>>>>>>>>>>>>>>>>>>>>

The  optical response of LFO with the radiation field {\bf E} $\perp$ {\bf c}  has been investigated  by Xu {\it et al.}, \cite{Xu} together with other properties,   between  30 meV and 6.5 eV and from 4 to 540 K. At zero field the authors observe two main peaks in the near-infrared/visible range, which is dominated by electronic excitations. The band at 1.1 eV is assigned to  Fe$^{2+}$/Fe$^{3+}$ charge transfer, and that at about 3.5 eV to a superposition of O$-p \to$Fe-$d$ and O$-p \to$Lu$-s$ charge transfer. The edge of the 1.1 eV peak  creates the insulating optical gap at $\sim$ 0.5 eV. A third feature in the mid infrared, peaked at about 0.3 eV, is attributed by the authors to on-site Fe$^{3+}$ excitations.  Li {\it et al.} \cite{THz} have instead explored the sub-THz response of  LFO by time-domain spectroscopy,  observing at low $T$ two collective excitations that they  assigned to a central mode and to a soft mode of the ferroelectric phase.

In the present experiment we have investigated both the sub-THz and the infrared optical conductivity of  LuFe$_{2}$O$_{4}$ by Fourier-Transform spectroscopy. We performed reflectivity  measurements  on a large single crystal from 7 to 16000 cm$^{-1}$ (i. e., from about 1 meV to 2 eV), in the temperature range 10 - 500 K, with the radiation field {\bf E} polarized both  $\perp$ {\bf c} and  $\parallel$ {\bf c}. We could thus observe in detail the optical phonon spectrum of LuFe$_{2}$O$_{4}$, and the striking modifications that it undergoes when crossing $T_{CO}$. In the sub-THz spectral range we detected at low $T$ a peak at 10 cm$^{-1}$, which corresponds approximately to the central mode reported in Ref. \onlinecite{THz}. We propose for that spectral feature an alternative interpretation which is related to charge order rather than to ferroelectricity.  

\section{Experiment}

The measurements were performed on a large (8x3x2 mm$^{3}$) single crystal of LuFe$_{2}$O$_{4}$, grown by optical floating zone melting method in a flowing argon atmosphere, \cite{Sun09} and characterized by X-ray diffraction and Laue imaging at room temperature.
The magnetic susceptibility of the sample, as measured in zero-fied cooling, is shown in Fig. \ref{chi}. In addition to the sharp peak at $T_N$ =   230 K, also reported previously, \cite{THz} we observe a change of slope at about 150 K which may be related to the change in the magnetic correlation length   observed in neutron scattering experiments around 175 K.

%<<<<<<<<<<<<<<<<<<<<<<<<<<<<<<<< FIG 2 >>>>>>>>>>>>>>>>>>>>>>>>>>>>>>>>>>

\begin{figure}[!b]   \begin{center}  
\leavevmode    
\epsfxsize=8.6cm \epsfbox {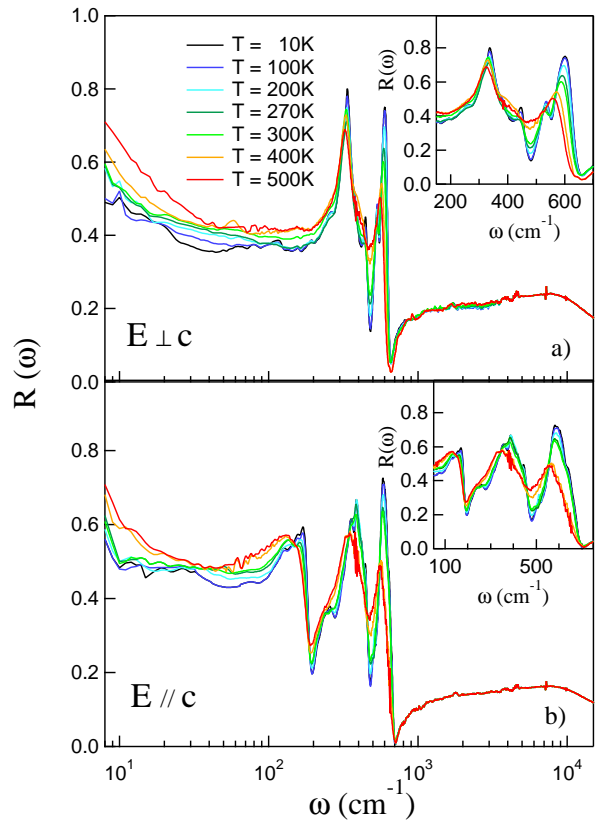}    
\caption{(Color online) Reflectivity of LuFe$_{2}$O$_{4}$ between 10K and 500K, as measured with the radiation field polarized along the $a$($b$)-axis (above)  and along the $c$-axis (below). The phonon region is reported in the inset on a linear frequency scale.}
\label{rif}
\end{center}
\end{figure}
%<<<<<<<<<<<<<<<<<<<<<<<<<<<<<<<< END FIG 2 >>>>>>>>>>>>>>>>>>>>>>>>>>>>>>>>>>

The largest crystal surface contained the axes $c$ and $a$ (or $b$). Its reflectivity $R(\omega)$ was measured between 7 and 16000 cm$^{-1}$  by Michelson interferometers, with the radiation field {\bf E} polarized either $\perp$ {\bf c} or  $\parallel$ {\bf c}.   Below 300 K the sample was theromoregulated within $\pm$ 2 K in a  He-flow cryostat, above 300 K in an evacuated capsule and in thermal contact with a heater. Both sample holders included a mobile hot filament for gold evaporation. Indeed,  after measuring the intensity $I_s (\omega)$ reflected by the sample, at each temperature gold was evaporated in situ, and the  intensity $I_0(\omega)$ reflected by the golden sample was measured for reference. The reflectivity $R(\omega) = I_s (\omega)/I_0(\omega)$ was thus obtained.  The sub-THz measurements were performed by the same procedure but using the coherent syncrotron radiation extracted from a bending magnet of the BESSY-II storage ring when working in its low-$\alpha$ mode. \cite{Abo} The real part of the optical conductivity was calculated by the use of Kramers-Kronig transformations. Standard extrapolations were applied to $R(\omega)$ at all temperatures,  at frequencies both lower and higher    than the measured interval.  

%<<<<<<<<<<<<<<<<<<<<<<<<<<<<<<<< FIG 3 >>>>>>>>>>>>>>>>>>>>>>>>>>>>>>>>>>

\begin{figure}[!t]   \begin{center}  
\leavevmode    
\epsfxsize=8.6cm \epsfbox {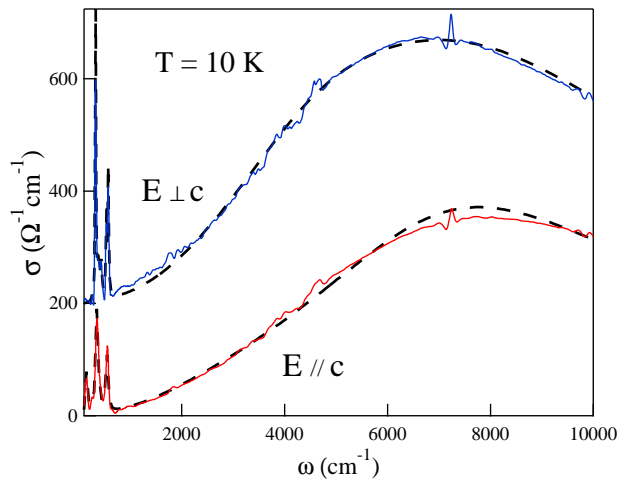}    
\caption{(Color online) Optical conductivity of LuFe$_{2}$O$_{4}$ at 10 K in the whole energy range, as measured with the radiation field  {\bf E}  $\perp$ {\bf c}  or {\bf E}  $\parallel$ {\bf c}. The fitting curves  are  reported by dashed lines. The zero of the upper curves has been shifted by 200 $\Omega^{-1}$cm$^{-1}$ for sake of clarity}
\label{sigma-all}
\end{center}
\end{figure}

%<<<<<<<<<<<<<<<<<<<<<<<<<<<<<<<< END FIG 3 >>>>>>>>>>>>>>>>>>>>>>>>>>>>>>>>>>

\section{Results and discussion} 

\subsection{Reflectivity and high-frequency conductivity} 

The reflectivity  of the LFO crystal is reported in Fig.\ref{rif} for {\bf E}  $\perp$ {\bf c} (a) and {\bf E}  $\parallel$ {\bf c} (b). In both panels $R(\omega)$ shows a rich and detailed optical-phonon spectrum, also reported  on a linear scale in the insets, which changes drastically at $T_{CO}$. In the {\bf E}  $\perp$ {\bf c} configuration,  at the lowest frequency and high $T$, the non-zero slope of $R(\omega)$ reveals a small and narrow Drude-like term, thermally activated.  Below $T_{CO}$, in the sub-THz region one instead detects  a weak but well defined peak at about 10 cm$^{-1}$. The mid- and the near-infrared ranges are finally occupied by a broad, practically $T$-independent band with a maximum around 8000 cm$^{-1}$ (1 eV). 

This broad peak is better seen in Fig. \ref{sigma-all}, which shows the real part of the optical conductivity, as extracted from $R(\omega)$ by the Kramers-Kronig transformations. The Figure reports the data at 10 K, but the high-frequency conductivity is virtually independent of temperature.  The best fit, also plotted in the Figure, requires a strong oscillator peaked at 7200 cm$^{-1}$ (0.9 eV) for {\bf E} $\perp$ {\bf c} and at 8400  cm$^{-1}$ (1.05 eV)  for {\bf E} $\parallel$ {\bf c}. The former one is in fair agreement with the contribution at 1.1 eV reported in Ref. \onlinecite{Xu} and assigned to the lowest-energy allowed electronic transition, \cite{Xiang} namely the Fe$^{2+}$ $\rightarrow$ Fe$^{3+}$  charge transfer. This may occur between one of the doubly degenerate $E$ levels, which slightly differ in energy from each other,  and the $A$-symmetry singlet at higher energy. \cite{Nagano}  

The authors of Ref. \onlinecite{Xu} also report a much weaker contribution at 0.3 eV,  detected in transmission measurements on thin samples.  They assign it to on-site Fe$^{3+}$ excitations, but it might also be due to the photoionization of the  polaronic charges in the CO state. \cite{Calvani} Indeed, the mobility of the charges in the CO regime is treated in Ref. \onlinecite{Xu} in terms of small-polaron hopping. In the present reflectivity spectra we do observe a weak, lower-energy sideband of the main peak, that however the fit places  (at 10 K) at 4000 cm$^{-1}$ (0.5 eV) for {\bf E} $\perp$ {\bf c} and at 4500  cm$^{-1}$ (0.55 eV)  for {\bf E} $\parallel$ {\bf c}.

\subsection{Optical conductivity in the phonon region} 

The high-temperature, rhombohedral $R\bar3m$ unit cell of LuFe$_{2}$O$_{4}$ contains 3 formula units, rotated by 120$^0$ around  {\bf c} with respect to one another, having the same 3 acoustic and 18 optical modes. The authors of Ref. \onlinecite{Harris} assumed  3 $E_{u}$ and 3  $E_{g}$ (doubly degenerate) optical phonons in the $ab$ plane, 3 $A_{2u}$ and  3 $A_{1g}$ (non-degenerate) optical modes for the $c$ axis, and calculated their frequencies  (the $u$ modes are infrared active, the $g$'s are Raman active). However, a factor group analysis \cite{Hateley} leads  us to a  more complex phonon scenario for LuFe$_{2}$O$_{4}$, as it is shown in Table I.

%<<<<<<<<<<<<<<<<<<<<<<<<<<<<<<<< TABLE I  >>>>>>>>>>>>>>>>>>>>>>>>>>>>>>>>>

\begin{table}[h]

\begin{center}
\begin{tabular}{{c}{c}{c}{c}}
\hline
\hline
\\

 Atoms 		 & Wyckoff 			& site 			 & Irreducible 		 \\ 
         			& notation 		& symmetry 		& representation 	 \\
 \\
 \hline
 \\
Fe, Fe		&	1(a), 1(b)		&	$D_{3d}$	&	2$E_u$+2 $A_{2u}$		 \\
Lu,O	 		&	2(c)			&	$C_{3v}$	&	$E_{g}+E_{u}$+$A_{1g}+A_{2u}$ \\	   
O,O,O  		&	3(d)			&	$C_{2h}$	&	$3E_{u}$+$A_{1u}+2A_{2u}$	       \\ 
\\
\hline
\hline

\end{tabular}
\end{center}
\label{symmetry}
\caption{Site symmetries of the 7 atoms per formula unit, and  irreducible representations of the vibrational modes, in the rhombohedral unit cell $R\bar3m$ of LuFe$_{2}$O$_{4}$.}
\end{table}

%<<<<<<<<<<<<<<<<<<<<<<<<<<<<<<<< END TABLE I  >>>>>>>>>>>>>>>>>>>>>>>>>>>>>>>>>

%<<<<<<<<<<<<<<<<<<<<<<<<<<<<<<<< TABLE II  >>>>>>>>>>>>>>>>>>>>>>>>>>>>>>>>>

\begin{table}[h]

\begin{center}
\begin{tabular}{{c}{c}{c}{c}{c}}
\hline
\hline
\\
		   & Infrared 	 &  Raman		& Acoustic		& Silent 	  \\ 
		    & active       	 &	active 		& 				& 		 \\
\\
\hline
\\
$ab$ plane & $5E_{u}$	&		$E_{g}$	 &	$E_{u}$		&		  \\
$c$ axis & $4A_{2u}$	&		$A_{1g}$	&	$A_{2u}$		&	$A_{1u}$   \\

\\
\hline
\hline
\end{tabular}
\end{center}
\label{modes}
\caption{Classification of the phonon modes predicted for the $ab$ plane and the $c$ axis, in the $R\bar3m$ high-temperature phase of  LuFe$_{2}$O$_{4}$.}
\end{table}

%<<<<<<<<<<<<<<<<<<<<<<<<<<<<<<<< END TABLE II >>>>>>>>>>>>>>>>>>>>>>>>>>>>>>>>

The resulting modes are classified in Table II with respect to their polarization and activity. In the  high-$T$ $R\bar3m$ symmetry,  the infrared-active modes are predicted to be  5 $E_{u}$ in the $ab$ plane, 4 $A_{2u}$ along  the $c$ axis. 
At low $T$, the CO reduces the crystal symmetry to the monoclinic $C2/m$, \cite{Angst}  which includes one site of symmetry $C_{2h}$(1), two $C_{i}$(2),  two $C_{2}$(2), and two $C_{s}$(2). Here, the number in parenthesis indicates how many atoms each site can accommodate. One thus obtains eight possible irreducible representations $\Gamma_i$, depending on the site occupation. The $\Gamma_i$'s with $i =1, ..., 4$ lead to 12 one-dimensional infrared-active modes, those with $i =5, ..., 8$ to 15 such modes. However, six out of them require that a site with inversion symmetry $C_{i}$ is occupied. As such an atom does not exist even in the high-$T$ phase, they can all  be excluded. The remaining two  irreducible representations of the crystal vibrations, in the low-$T$ $C2/m$ phase, are

\begin{eqnarray}
\Gamma_1=4A_u+8B_u+2A_g+4B_g \nonumber \\
\Gamma_2=4A_u+8B_u+3A_g+3B_g
\label{c2/m}
\end{eqnarray}

\noindent
They  differ for the Raman spectrum but predict the same 12 infrared-active, non-degenerate, modes. One thus obtains a  reliable prediction on the effect of the symmetry reduction below $T_{CO}$ on the phonon spectrum. 

%<<<<<<<<<<<<<<<<<<<<<<<<<<<<<<<< FIG 4 >>>>>>>>>>>>>>>>>>>>>>>>>>>>>>>>>>

\begin{figure}[!t]   \begin{center}  
\leavevmode    
\epsfxsize=8.6cm \epsfbox {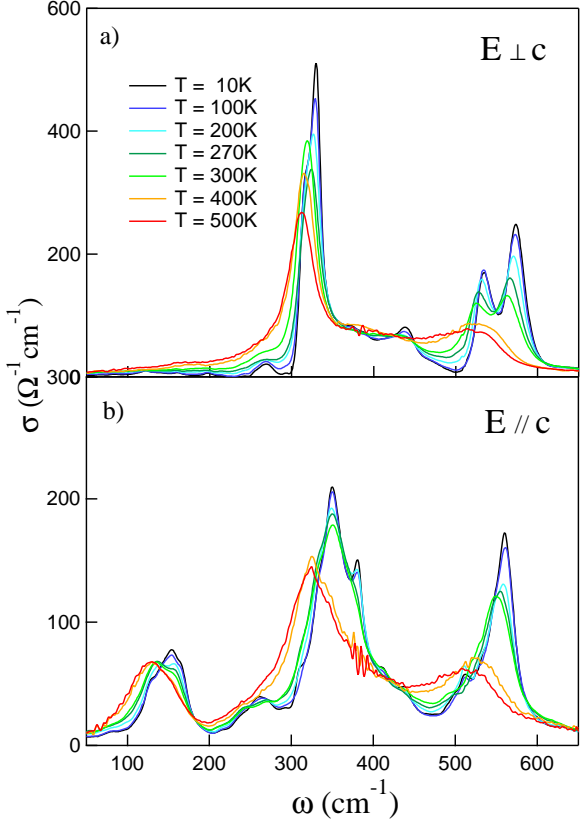}    
\caption{(Color online) Optical conductivity of LuFe$_{2}$O$_{4}$ between 10K and 500K in the phonon region, as measured with the radiation field  {\bf E}  $\perp$ {\bf c}  and {\bf E}  $\parallel$ {\bf c}.}
\label{sig}
\end{center}
\end{figure}

%<<<<<<<<<<<<<<<<<<<<<<<<<<<<<<<< END FIG 4 >>>>>>>>>>>>>>>>>>>>>>>>>>>>>>>>>>

This effect is clearly seen in  Fig.\ref{sig}, which shows the optical conductivity of LFO measured in the far infrared.  Below 320 K, for {\bf E}  $\perp$ {\bf c}  the broad contribution at  530 cm$^{-1}$ splits into two components separated by about 40 cm$^{-1}$;  for {\bf E}  $\parallel$ {\bf c} all lines shift abruptly to higher frequencies and new components do appear. Incidentally, no phonon line displays Fano-like asymmetries, \cite{Lupi98} thus confirming the absence of a free-carrier background in the phonon energy range. 

In order to obtain a closer comparison with theory, the optical conductivity was fit, through the relation $\sigma = (\omega /4\pi) \epsilon_{2}$, to the imaginary part $\epsilon_{2}$ of the Lorentzian dielectric function   

\begin{equation}
\tilde \epsilon (\omega)=\epsilon_{1}+i \epsilon_2 =\epsilon_{\infty}+\sum_{j=1}^{n}\frac{A_{j}\Omega^{2}_{TO_{j}}}{\Omega^{2}_{TO_{j}}-\omega^{2}-i\gamma_{j}\omega} \\
\label{epsilon}
\end{equation}

\begin{figure}[t]
\begin{center}
\leavevmode    
\epsfxsize=8.6cm 
\epsfbox{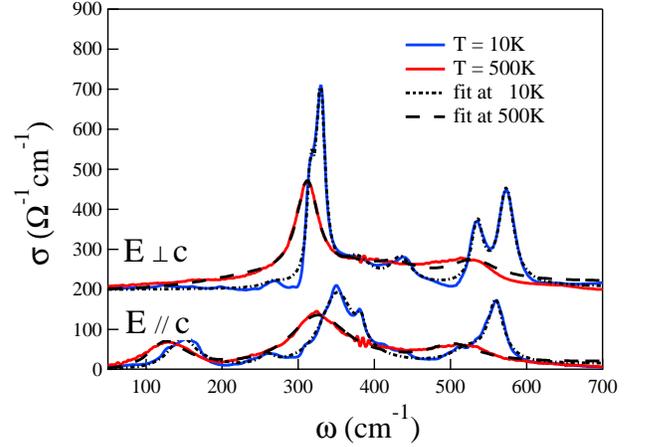}
\caption{Optical conductivity  at 10K and 500K, for both {\bf E}  $\perp$ {\bf c}  and {\bf E}  $\parallel$ {\bf c}, with the fitting curves which provide the parameters listed in Table III. The zero of the upper curves has been shifted by 200 $\Omega^{-1}$cm$^{-1}$ for sake of clarity.}
\label{s1_phon_fit}
\end{center}
\end{figure}

\noindent
Therein, $\epsilon_{\infty}$ accounts for the high-frequency contributions to $\tilde \epsilon(\omega)$ , while  $\Omega_{TO_{j}}$, $A_{j}$ and $\gamma_{j}$ are the central frequency,  strength and damping of the $\emph{j}$-th mode, respectively. 
The accuracy of the procedure can be evaluated in Fig. \ref{s1_phon_fit}, where the fitting curves are compared with the experimental $\sigma (\omega)$ at both extreme temperatures. 

The parameters  obtained by fitting Eq.  \ref{epsilon} to the optical conductivity of Fig. \ref{sig} with the minimum possible number of oscillators,  both in the rhombohedral $R\bar3m$ cell at 500 K, and in the monoclinic $C2/m$ cell at 10 K, are listed in Table III.  Therein, a  good agreement with the factor-group predictions reported  in Table II and in Eq. \ref{c2/m} is displayed. At high $T$ one obtains a good fit by using  5  Lorentzians in the $ab$ plane,  3  along the $c$ axis, to be compared with the 5 and 4 infrared modes, respectively, predicted in Table II. At low $T$, in the monoclinic symmetry, the best fit requires 6 modes for {\bf E}  $\perp$ {\bf c}   and 6 for {\bf E}  $\parallel$ {\bf c} , in excellent agreement with the  12 infrared active phonons predicted in total by Eq. \ref{c2/m}. A shoulder which appears at 316 cm$^{-1}$ both for {\bf E}  $\perp$ {\bf c}  and {\bf E}  $\parallel$ {\bf c} is unlikely to be due to a normal mode of the crystal.

\begin{figure}[t]
\begin{center}
\leavevmode    
\epsfxsize=8.6cm 
\epsfbox{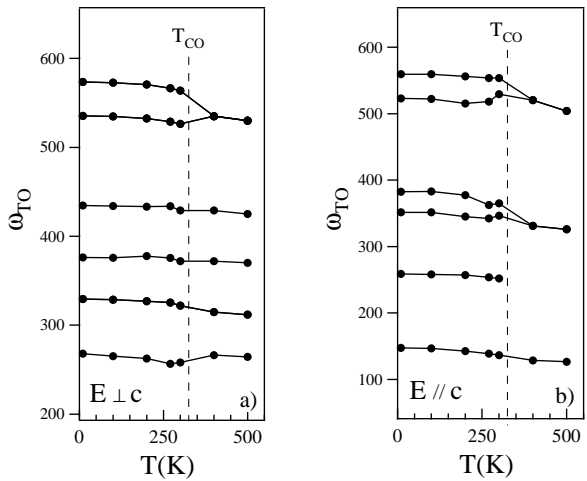}
\caption{Temperature dependence of the  phonon frequencies for  {\bf E}  $\perp$ {\bf c} (left)  and {\bf E}  $\parallel$ {\bf c} (right) in LuFe$_{2}$O$_{4}$.}
\label{frequencies}
\end{center}
\end{figure}

The temperature evolution of the main phonon frequencies is shown in Fig. \ref{frequencies} for both the $ab$ plane and the $c$ axis. Therein one may clearly see the branching at $T_{CO}$ of the highest-energy mode  of the $ab$ plane and of two modes of the $c$ axis. In all these cases,  a narrower line originates from the high-$T$ absorption at a moderately higher frequency, while a new mode springs up at a much higher frequency.

%<<<<<<<<<<<<<<<<<<<<<<<<<<<<<<<< FIG 5 >>>>>>>>>>>>>>>>>>>>>>>>>>>>>>>>>>

%<<<<<<<<<<<<<<<<<<<<<<<<<<<<<<<< END FIG 5 >>>>>>>>>>>>>>>>>>>>>>>>>>>>>>>>>>

%<<<<<<<<<<<<<<<<<<<<<<<<<<<<<<<< FIG 6 >>>>>>>>>>>>>>>>>>>>>>>>>>>>>>>>>>>>>

%<<<<<<<<<<<<<<<<<<<<<<<<<<<<<<<< END FIG 6 >>>>>>>>>>>>>>>>>>>>>>>>>>>>>>>>>>

%<<<<<<<<<<<<<<<<<<<<<<<<<<<<<<<< TABLE III >>>>>>>>>>>>>>>>>>>>>>>>>>>>>>>>>>

\begin{table}[h]

\begin{center}
\begin{tabular}{{l}{c}{c}{c}{c}{c}{c}{c}{c}}
\hline
\hline
\\

 Mode&$\Omega_{TO}$&$\gamma$ & $A$ & $\Omega_{calc}$ &$\Omega_{TO}$& $\gamma$ & $A$ \\ 
 \\
 \hline
 &&&&{\bf E}  $\perp$ {\bf c}&&&\\  
& &10K& & &&500K &\\ 
 \hline
\\
	&		&		&		&	92	&		&		&		\\
a1	&	268	&	30	&	150	&		&	264	&	34	&	190	\\
	&	316	&		&		&		&		&		&		\\
a2	&	329	&	14	&	642	&	332	&	312	&	35	&	700	\\
a3	&	376	&	50	&	441	&		&	370	&	43	&	215	\\
a4	&	435	&	34	&	361	&		&	425	&	85	&	441	\\
a5	&	535	&	18	&	408	&	474	&	530	&	75	&	473	\\
a6	&	573	&	23	&	579	&		&		&		&		\\
\\
 \hline
&&&&{\bf E}  $\parallel$ {\bf c}&&&\\
&& 10K& & &&500K\\
\hline

c1 	&	148	&	46	&	461	&	161	&	127	&	59	&	493	\\
c2    	&	259	&	22	&	175	&		&		&		&		\\
	&	316	&		&		&		&		&		&		\\
c3  	&	352	&	56	&	787	&	310	&	326	&	95	&	856	\\
c4   	&	383	&	12	&	196	&		&		&		&		\\
c5  	&	523	&	55	&	341	&	465	&	504	&	83	&	466	\\
c6    	&	559	&	24	&	467	&		&		&		&		\\

\\
\hline
\hline
\end{tabular}
\end{center}
\label{param_fit}
\caption{Parameters obtained by fitting Eq.  \ref{epsilon} to the phonon spectrum of LuFe$_{2}$O$_{4}$, as measured with the radiation field polarized in the $ab$-plane and along the $c$-axis. The calculated values are those reported in Ref. \onlinecite{Harris}. All figures are in cm$^{-1}$. }
\end{table}

%<<<<<<<<<<<<<<<<<<<<<<<<<<<<<<<< END TABLE III >>>>>>>>>>>>>>>>>>>>>>>>>>>>>>>>>>

\subsection{Optical response in the sub-THz region} 

\begin{figure}[t]
\begin{center}
\leavevmode    
\epsfxsize=8.6cm 
\epsfbox{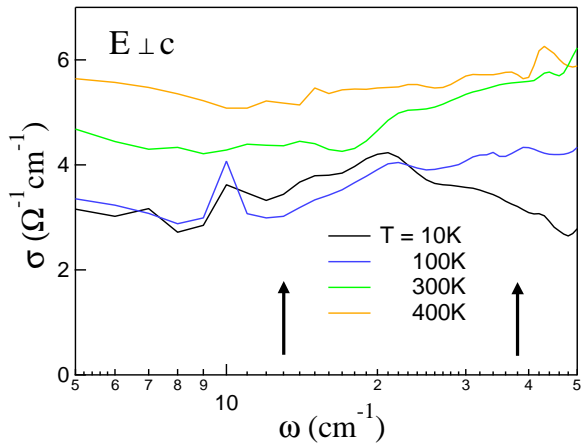}
\caption{Optical conductivity  for {\bf E}  $\perp$ {\bf c} of LuFe$_{2}$O$_{4}$ in the sub-THz region at different temperatures. The arrows mark the frequencies of the peaks in $\epsilon_2$ extracted from a time-domain experiment. \cite{THz} }
\label{subTHz}
\end{center}
\end{figure}

As already mentioned, we have extended our investigation well below the phonon region, in the sub-THz range, by use of coherent  synchrotron radiation. We could thus greatly enhance the signal-to-noise ratio on the reflectivity, while maintaining a low perturbation on the system. The $R(\omega)$ thus obtained can be seen in Fig. \ref{rif}  thanks to its logarithmic frequency scale. Therein, at 100 and 10 K  a well defined peak appears at $\omega_0$ = 10 cm$^{-1}$. This extremely low-frequency feature is then associated with  the three-dimensional charge order, further reinforced by the ferrimagnetic transition at 240 K.  The resulting sub-THz optical conductivity is reported in Fig. \ref{subTHz}, with the  feature at $\omega_0$  again evident at low temperature. In Ref. \onlinecite{THz}, a peak was detected in the $\epsilon_2$ extracted from time-resolved sub-THz spectroscopy. \cite{THz}   At 50 K it was reported at 13 cm$^{-1}$ (left arrow in Fig. \ref{subTHz}), in agreement within errors with the present $\omega_0$. The authors of Ref. \onlinecite{THz} attributed their peak to a central mode of the ferroelectric LFO phase. They observed a further, broad contribution around  40 cm$^{-1}$ (right arrow in Fig. \ref{subTHz}), which was assigned to a soft mode. In Fig. \ref{subTHz}, at 10 K, a second absorption possibly appears  between  20 and 30 cm$^{-1}$.

%<<<<<<<<<<<<<<<<<<<<<<<<<<<<<<<< FIG 7 >>>>>>>>>>>>>>>>>>>>>>>>>>>>>>>>>>>>>>

%<<<<<<<<<<<<<<<<<<<<<<<<<<<<<<<< END FIG 7 >>>>>>>>>>>>>>>>>>>>>>>>>>>>>>>>>>

The present observations may be consistent with those of time-resolved spectroscopy,  if one considers that   in polycrystalline samples, as is the case of Ref. \onlinecite{THz} , the soft mode stiffens with respect to that observed in single crystals. \cite{Petzelt} 
However, if one considers that LuFe$_{2}$O$_{4}$, as already mentioned, does not exhibit bulk ferroelectricity at zero field, there are alternative explanations for the present peak at 10 cm$^{-1}$ (and for the broad band at higher frequency, if confirmed by further experiments). Indeed, similar features have been observed in several systems where the charge order can be described in terms of charge density waves (CDW) and this is also the case of LFO. \cite{Yamada} For example, in the  manganite La$_{0.25}$Ca$_{0.75}$MnO$_3$, below the charge ordering temperature $T_{CO}$ a narrow peak at 7.5 cm$^{-1}$ is observed, followed by a broad absorption at $\sim$ 30 cm$^{-1}$. \cite{Nucara} The former feature has been interpreted as a collective excitation (phason) of the CDW, which displaces from zero to a finite frequency when either it is pinned to lattice impurities or  the charge order is commensurate with the lattice. The broader band was ascribed to a combination band of the phason with the amplitudon, another excitation which is present in a CDW. \cite{Nucara} According to other authors, the highest-frequency feature is instead due to an acoustic phonon, which becomes infrared active at zero wavevector due to the folding of the Brillouin zone determined by the CO. \cite{Dressel} Further  sub-THz experiments are probably needed, before reaching a common interpretation of the elusive excitations which are detected at those low frequencies in the charge-ordered systems. 

\section{Conclusion}

In conclusion, we have reported here a study of the charge-ordered multiferroic LuFe$_{2}$O$_{4}$ in the  infrared and the sub-THz range, with special focus on the optical phonon region. Therein we have observed the dramatic effect of the symmetry reduction - caused by the transition at $T_{CO}$ = 320 K between the 2-D and 3-D charge order -  both on the modes of the $ab$ plane and on those of the $c$ axis of the rhombohedral structure. Below $T_{CO}$ some phonon bands split into pairs of narrower lines, while new modes do appear.  The number of the observed phonon lines, both above and below $T_{CO}$, is in good agreement with a factor-group analysis of the respective cell symmetries. No evident effects on the phonon spectrum is instead detected when crossing the ferrimagnetic transition at 240 K.  In the sub-THz region, a very weak electric conduction, thermally activated, is detected at high temperature, while a peak  becomes evident  in the low-temperature optical conductivity, possibly accompanied by a broad absorption at higher frequency. For that feature, similar to the one reported previously in time-domain experiments, we have proposed an interpretation related to the charge order rather than to ferroelectricity, and based on a comparison with similar sub-THz spectra recently obtained in charge-ordered manganites at low temperature.
%<<<<<<<<<<<<<<<<<<<<<<<<<<<<<<<< BIBLIOGRAPHY >>>>>>>>>>>>>>>>>>>>>>>>>>>>>>>>>

%<<<<<<<<<<<<<<<<<<<<<<<<<<<<<<<< END BIBLIOGRAPHY >>>>>>>>>>>>>>>>>>>>>>>>>>>>>

\end{document}